\documentclass[10pt,journal,compsoc]{IEEEtran}
\usepackage{graphicx}
\usepackage{booktabs}
\usepackage{color}
\usepackage{array}
\usepackage{hyperref}

\newif\ifshowchanges
\showchangesfalse
\newcommand{\newchange}[1]{\ifshowchanges{\color{blue}#1}\else#1\fi}

\begin{document}

\title{The Biggest Risk of Embodied AI\\
is Governance Lag}

\author{
    Shaoshan~Liu
}

\maketitle
\begin{abstract}
Embodied AI is widely discussed as a job-displacement problem. The deeper risk, however, is governance lag: the time and capability gap between a measurable change in technology deployment and an institutional response able to address its consequences. Building on the established pacing problem and the Collingridge dilemma, this article argues that embodied AI intensifies that gap through scalable models and platforms, task-level reorganization, and the separation of upstream technological control from downstream social impact. We distinguish three mutually reinforcing forms of lag, observational, institutional, and distributive, and propose a compliance architecture based on deployment visibility, stack-level accountability, trigger-based adjustment, and automatic distributional response. The central policy challenge is not automation alone, but whether governance systems can become observable, responsive, and adaptive before disruption becomes entrenched.
\end{abstract}

\section{Introduction}
\label{sec:intro}

Embodied AI refers to intelligent systems integrated with physical bodies that can perceive, decide, and act in the real world. It includes industrial robots, autonomous mobile systems, humanoids, and other machines that combine AI with sensing, motion, and task execution~\cite{liu2026evolutionary}. As reusable robotic platforms are increasingly paired with more general AI models, embodied AI is beginning to move beyond the traditional robotics paradigm of highly customized machines built for narrow tasks. This development points toward an emerging autonomy economy in which improvements in models, software, data, and deployment infrastructure could be reused across applications and create scale effects beyond any single machine or use case~\cite{liu2024shaping,wu2025autonomy}.

Most public debate, however, frames embodied AI primarily as a labor problem. The dominant concern is that smarter robots will replace large numbers of human workers across factories, warehouses, transport systems, and care settings. That concern is not misplaced. Industrial robotics combines physical task substitution, capital deepening, and workplace reorganization at scale.

Deployment is also accelerating, although the available large-scale evidence still comes primarily from conventional industrial robotics rather than from general-model embodied AI. In 2024, manufacturing robot density reached 177 units per 10{,}000 employees worldwide and 204 in Asia, while China accounted for 295{,}045 new industrial-robot installations, representing 54\% of global demand~\cite{worldrobotics2025industrial,ifr2025china2million}. These figures should be read as a leading indicator of the physical automation base into which more general embodied-AI capabilities may diffuse, not as evidence that general-model systems are already deployed at comparable scale. Direct deployment data for reusable, general-model embodied systems remain thin. That limitation does not establish rapid diffusion; it underscores why a dedicated framework for observing deployment and its effects is needed.

But job replacement is not the core problem brought by embodied AI. The deeper risk is governance lag: the widening gap between the speed of embodied AI deployment and the much slower pace at which public institutions can observe, interpret, and respond to its effects. Once embodied AI is understood as part of the autonomy economy, the policy challenge becomes broader than automation alone. The issue is not only whether individual systems can perform specific tasks, but whether governments can keep pace with a technology that may reorganize work, reshape firm structure, and concentrate gains faster than existing institutions can adapt~\cite{liu2024shaping,liu2025human}.

The general problem of technology outpacing governance is well established. The ``pacing problem'' describes the growing gap between rapid technological change and slower legal and institutional adaptation~\cite{marchant2011growing}. The Collingridge dilemma identifies a related trade-off: emerging technologies are easiest to shape before their consequences are fully known, but become harder to control after their effects are visible and the technology is deeply embedded~\cite{collingridge1980social}. This article therefore does not present governance lag as a wholly new phenomenon. Its contribution is to show how the pacing problem takes a distinctive form in embodied AI and to decompose that form into three analytically useful lags.

In this article, \textbf{we argue that embodied AI should be understood not only as an automation challenge, but as a governance challenge.} The issue is not simply whether jobs will be displaced, but whether states can build governance and compliance systems fast enough to match how embodied AI is deployed and how its effects unfold. If not, the greater danger will not be automation alone, but a gap in public visibility, institutional adjustment, and timely response.

Calling governance lag the central risk does not claim that it will always produce more direct harm than job displacement. The claim is instead that governance lag is upstream: it determines whether displacement, concentration, safety risks, and distributional pressures can be detected and mitigated before they become entrenched. In that sense, it is a risk multiplier rather than a substitute for the underlying harms.

\begin{figure*}[t]
\centering
\includegraphics[width=\textwidth]{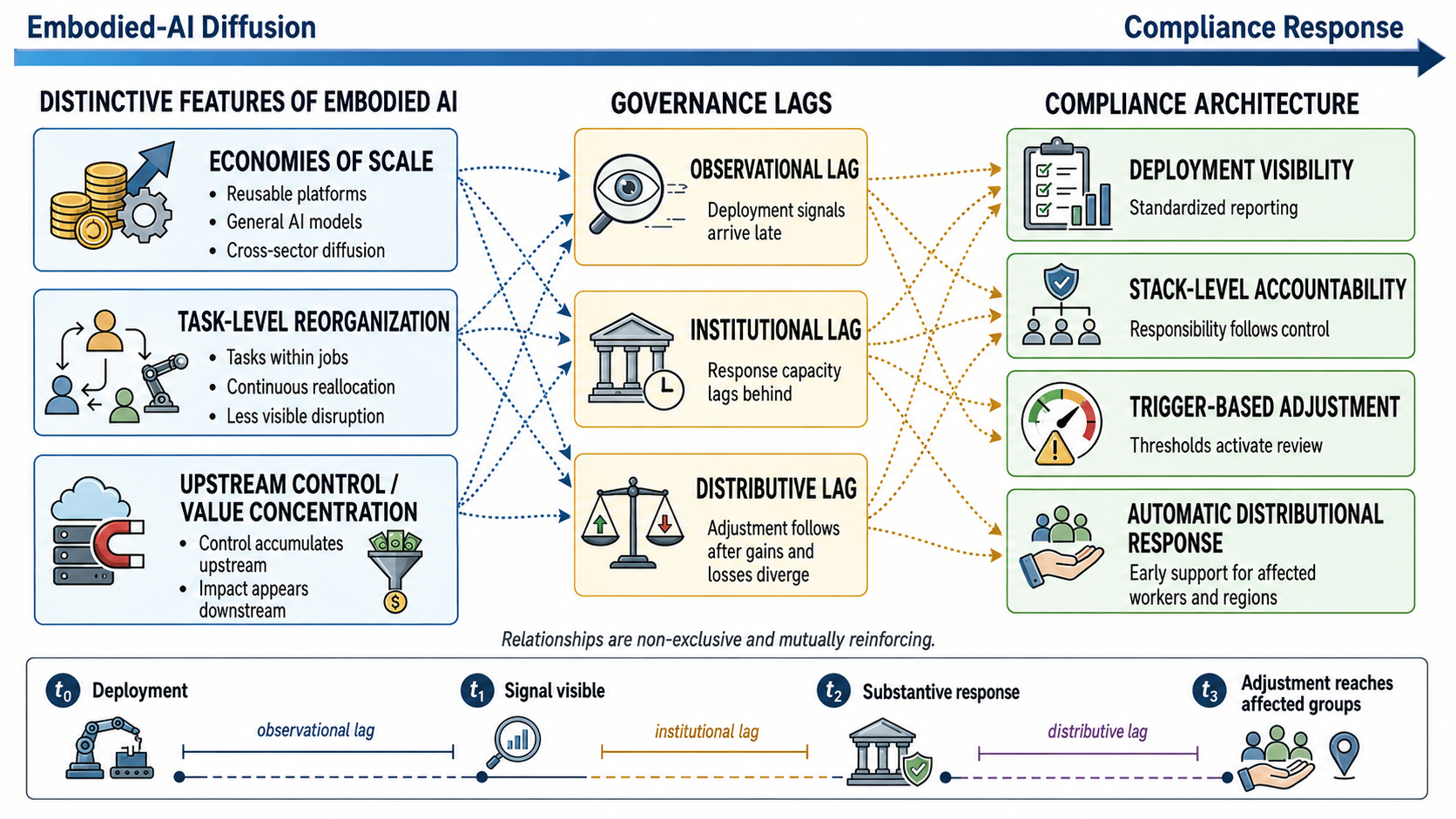}
\caption{Analytical framework linking embodied-AI diffusion to three mutually
reinforcing governance lags and a proposed compliance architecture. Dotted
arrows indicate non-exclusive relationships. The lower timeline illustrates a
possible, rather than universal, timing asymmetry between deployment,
visibility, institutional response, and distributional adjustment.}
\label{fig:overview}
\end{figure*}

Figure~\ref{fig:overview} summarizes the paper's analytical framework, linking the diffusion of embodied AI across the physical economy to its distinctive governance challenges and the corresponding compliance response.

\section{Embodied AI as a Governance Problem}
\label{sec:gov}

If the central risk of embodied AI is governance lag, then the next question is why that lag matters. The answer is historical as much as technological. Periods of rapid change become destabilizing not simply when production changes, but when institutions fail to adjust at the same speed.

Historical cases illustrate this relationship, but they do not establish a simple causal path from technological change to political breakdown. In late Tsarist Russia, rapid industrialization interacted with weak representation, repression, war, and pre-existing political fragility; in Weimar Germany, mass unemployment interacted with fiscal constraints, social conflict, and an already vulnerable political order~\cite{figes1996peoplestragedy,james1986germanslump}. Unemployment rose from just under 1.3 million in summer 1929 to more than 6 million by early 1932, but institutional lag was one contributing factor among several rather than a sufficient explanation. These cases are therefore illustrative rather than predictive. \newchange{They were selected because they provide clear historical examples of rapid economic transformation interacting with institutional weakness; they are not intended to imply that institutional failure is the modal outcome of technological change.} Their narrower lesson is that the consequences of a production shock depend partly on whether institutions can observe disruption, coordinate adjustment, and support affected groups.

Embodied AI may raise this problem in a more acute form than earlier waves of mechanization. Like prior automation technologies, it can substitute for labor, reorganize production, and concentrate gains. But it also differs from earlier industrial transformations in three important ways.

First, embodied AI is increasingly being built on reusable robotic platforms linked to general AI models, giving it the potential to achieve economies of scale in a way earlier automation systems generally could not. Traditional industrial automation was usually tied to highly specific machines and narrowly engineered workflows, which meant that improvements were largely local and application-specific. By contrast, recent embodied-AI systems are moving toward more general-purpose architectures in which the same underlying models can support planning, perception, and control across multiple tasks and embodiments~\cite{driess2023palm,reed2022generalist}. As a result, advances in models, software, and data could be reused across deployments, reducing some marginal adaptation costs once physical integration constraints are overcome.

The qualification is important: these architectures remain largely at the research, pilot, or early-commercial stage, so economy-wide scale effects are a structural potential rather than an established empirical outcome. Diffusion may also proceed more slowly than model capability alone suggests because physical deployment requires capital expenditure, safety certification, site integration, maintenance, and workflow redesign. Their governance significance lies in the possibility that once these physical constraints are overcome and a general capability crosses a deployment threshold, improvements can propagate through software and platform updates faster than sector-specific institutions can adjust.

Second, embodied AI creates a sharper governance problem because it can reallocate tasks before it eliminates jobs. Acemoglu and Restrepo's framework shows that automation changes labor demand by shifting tasks from labor to capital, not only by removing entire occupations~\cite{acemoglu2019automation}. The measurement problem is concrete: employment and unemployment statistics record whether a worker holds a job, but generally do not reveal whether the worker's task mix, autonomy, pace, or hours attached to particular activities have changed. Task-level reallocation can therefore precede any visible change in occupation-level employment.

In a warehouse, for example, autonomous mobile robots may take over internal transport while workers continue picking, packing, exception handling, and quality checks. The occupation remains, but its task composition and labor demand change incrementally. Such changes may first appear through wage pressure, hours changes, workflow control, or local labor-market effects rather than through a discrete mass-layoff event~\cite{acemoglu2019automation,acemoglu2020robots}. Public labor-market and social-protection systems are generally better suited to visible employment shocks than to this gradual reallocation, and coverage remains incomplete even for conventional shocks~\cite{worldbank2025statesocialprotection}.

Third, embodied AI creates a sharper governance problem because it has structural tendencies toward concentration. The reason is that embodied AI depends on a vertically layered technology stack spanning compute, models, robotic hardware, systems integration, and deployment channels. In such a system, control can accumulate upstream even when downstream deployment remains fragmented. OECD analysis shows that AI infrastructure markets can exhibit high concentration, barriers to entry, switching costs, and vertical relationships that reinforce market power~\cite{oecd2025aiinfrastructure}. The CMA likewise warns that control over foundation models and related partnerships can shape downstream competition~\cite{cma2024foundationmodels}.

Layering alone is not new: traditional automation also relied on manufacturers, integrators, suppliers, and operators. The distinctive issue is the separation between the locus of control and the locus of social impact. A cloud, model, or platform provider may shape capability, interoperability, switching costs, and access upstream, while employment, safety, and regional effects materialize downstream at deployment sites. Accountability and market power can therefore diverge from the places where social consequences are observed~\cite{oecd2025aiinfrastructure,cma2024foundationmodels,liu2024shaping}. This separation makes concentrated control harder to identify and responsibility harder to assign through institutions designed around a single firm, product, or sector.

\begin{table}[t]
\caption{Why embodied AI creates a sharper governance problem than earlier automation waves}
\label{tab:embodied_vs_automation}
\centering
\footnotesize
\begin{tabular}{p{0.14\linewidth} p{0.23\linewidth} p{0.23\linewidth} p{0.26\linewidth}}
\toprule
\textbf{Dimension} & \textbf{Earlier automation} & \textbf{Embodied AI} & \textbf{Governance implication} \\
\midrule
Scale & Local, fixed systems & Reusable models and platforms & Potentially faster cross-sector diffusion \\
\addlinespace
Substitution & Jobs, stages & Tasks within jobs & Disruption is less visible \\
\addlinespace
Control & Control near deployment & Upstream control, downstream impact & Responsibility and market power diverge \\
\bottomrule
\end{tabular}
\end{table}

Table~\ref{tab:embodied_vs_automation} summarizes why embodied AI creates a sharper governance challenge than earlier automation waves. That is why embodied AI must be treated first as a governance problem. The central question is not only what the technology can do, but whether states can build the visibility, compliance, and adjustment capacity needed to govern it before disruption becomes entrenched.

\section{The Three Lags of Embodied AI Governance}
\label{sec:lag}

If the problem of embodied AI is ultimately one of governance, then the next question is how that governance challenge can be detected and assessed. We distinguish three lags: observational, institutional, and distributive. They are analytically distinct and mutually reinforcing. They often unfold in sequence, limited visibility weakens institutional response, which can worsen distributional outcomes, but the ordering is not universal. Distributional pressures can emerge even when deployment is well observed, and institutional mismatch may persist despite adequate data.

For analytical purposes, governance lag can be understood as the delay and capability mismatch between a measurable change in embodied-AI deployment and an institutional response able to address its consequences. At the sector level, it can be assessed through three practical questions: how long it takes for deployment and task-level effects to become reliably visible; how long institutions take to initiate a substantive response after a signal is verified; and how long it takes for adjustment support to reach affected workers or regions. A response is substantive when it changes an enforceable obligation, initiates a formal regulatory or competition review, activates a funded adjustment mechanism, or delivers material support. Public statements and preliminary consultations alone do not end the lag.

These time intervals should be considered together with coverage and capability indicators. Observational lag can be assessed through reporting frequency, reporting latency, and the share of deployments covered by standardized disclosure. Institutional lag can be assessed through the delay between a verified signal and regulatory review, retraining provision, competition intervention, or another relevant action. Distributive lag can be assessed through the delay between measurable effects on wages, hours, or regional employment and the delivery of adjustment support. A sector may therefore respond quickly to the deployments it can see while still exhibiting substantial governance lag if reporting covers only a small share of deployments or omits task-level effects.

The first is \textit{observational lag}. No widely adopted public framework yet tracks embodied-AI deployment and its labor-market effects with sufficient speed and precision. Existing statistics capture fragments of the picture, but not the phenomenon as a coherent policy object. Industrial-robotics statistics provide one partial view of physical automation ~\cite{worldrobotics2025industrial}, and China's 2026 national standard system for humanoid robotics and embodied intelligence reflects early official recognition of the field~\cite{miit2026humanoidstandards}. But neither provides what governance actually requires: timely visibility into where embodied AI is being deployed, which tasks it is substituting, how work is being reorganized, and how wages, hours, and employment conditions are changing as a result.

The limitation is specific. Robot-density statistics measure the operational stock of industrial robots per 10{,}000 manufacturing employees, but they do not identify whether systems use reusable general models, which tasks are transferred, or how work changes after deployment. China's 2026 standard-system initiative defines priorities across six areas of humanoid robotics and embodied intelligence, but it does not itself establish recurring deployment-level or labor-effect reporting ~\cite{miit2026humanoidstandards}. Technical definition and safety classification are essential foundations, yet they are not substitutes for recurrent observability.

The second is \textit{institutional lag}. Even when governments can see the shock, their core institutions remain poorly matched to the mechanisms through which embodied AI changes production. Labor institutions are still organized mainly around jobs and firms, while embodied AI often substitutes tasks within jobs and allows firms to reorganize work incrementally rather than all at once~\cite{acemoglu2019automation}. Competition institutions are also misaligned. Embodied AI is not delivered through a single product market, but through a layered stack of compute, models, robot platforms, systems integration, and deployment channels. As control over these layers becomes more concentrated, the sources of market power become harder to see through conventional sectoral boundaries~\cite{oecd2025aiinfrastructure}. Training institutions lag as well. Firms can alter workflows quickly, but worker retraining and occupational adjustment move much more slowly. The result is a widening mismatch between the speed of technological deployment and the speed of institutional adaptation.

The third is \textit{distributive lag}. Embodied AI may raise productivity quickly, but redistribution and adjustment systems are usually designed to respond slowly and after the fact. Gains often accrue first to the owners of platforms, models, deployment channels, and capital-intensive production systems, while losses spread through wages, work organization, bargaining power, and local labor markets. Existing social-protection systems are poorly matched to this pattern because they often respond after formal job loss and have incomplete coverage~\cite{worldbank2025statesocialprotection}. This creates an asymmetry: concentration can happen early, while adjustment comes late. When that gap widens, economic change becomes harder to absorb socially and politically.

Governance lag does not imply a regulatory vacuum. Existing instruments already address important parts of the problem, but they differ materially in legal status, scope, and governance function. Table~\ref{tab:existing_instruments} compares the principal instruments discussed here and identifies the functions that remain outside their current scope.

\begin{table*}[t]
\caption{Existing instruments and the remaining embodied-AI governance gaps}
\label{tab:existing_instruments}
\centering
\footnotesize
\setlength{\tabcolsep}{4pt}
\renewcommand{\arraystretch}{1.12}
\begin{tabular}{
    p{0.22\textwidth}
    p{0.16\textwidth}
    p{0.24\textwidth}
    p{0.29\textwidth}}
\toprule
\textbf{Instrument} &
\textbf{Status} &
\textbf{Primary coverage} &
\textbf{Key remaining gap} \\
\midrule

China's 2026 standard system
~\cite{miit2026humanoidstandards}
&
National standardization framework
&
Technical and industrial standardization priorities
&
No recurring deployment or labor-effect reporting \\

\addlinespace

EU AI Act, as amended in 2026
~\cite{euaiact2024,eudigitalomnibus2026}
&
\newchange{Binding legislation; staged application}
&
Risk-based obligations for AI systems
&
No economy-wide task-level labor observability \\

\addlinespace

EU Machinery Regulation and robot-safety standards
~\cite{eumachinery2023,iso10218_1_2025,iso10218_2_2025,iso15066_2016}
&
\newchange{Binding legislation and technical standards; staged application}
&
Machinery, robot, cell, and collaborative-operation safety
&
No cross-stack social-impact accountability or adjustment triggers \\

\addlinespace

ISO/IEC 42001:2023
~\cite{iso42001_2023}
&
Management-system standard
&
Organizational AI governance processes
&
No public deployment visibility or economy-wide intervention mechanism \\

\addlinespace

NIST AI RMF
~\cite{nistairmf2023}
&
Voluntary framework
&
Lifecycle AI risk identification and management
&
No mandatory disclosure, enforcement, or distributional response \\

\bottomrule
\end{tabular}
\vspace{2pt}

\newchange{\parbox{0.94\textwidth}{\scriptsize \textit{Effective-date staging:}
Under the consolidated EU texts, the AI Act applies generally from 2 August 2026, but Chapter III, Sections 1--3 apply from 2 December 2027 for high-risk systems classified under Article 6(2) and Annex III, and from 2 August 2028 for high-risk systems classified under Article 6(1) and Annex I. The Machinery Regulation applies generally from 20 January 2027, with specified provisions applying earlier under Article 54.}}
\end{table*}

The comparison reveals that the remaining problem is not an absence of governance, but a mismatch among governance functions. Binding legislation primarily governs defined system and product risks. Robot-safety standards address physical hazards and system integration. Management-system standards and voluntary frameworks support organizational risk processes. These instruments are important, but they do not collectively produce recurrent deployment visibility, task-level labor observability, responsibility across
the embodied-AI stack, or timely distributional adjustment.

The gap is therefore one of fragmentation across legal, technical, labor, competition, and social-protection regimes. \newchange{The proposed compliance architecture is intended to connect governance functions across these heterogeneous instruments and regimes, supplementing rather than displacing binding legislation, technical and management-system standards, and voluntary guidance that already apply or are scheduled to apply.}

These three lags are analytically distinct, but they reinforce one another. Observational lag delays recognition. Institutional lag weakens response. Distributive lag magnifies the consequences of both, while also being capable of arising independently. Together, they explain why the main risk of embodied AI is not simply that jobs may be replaced, but that governance systems may fall behind a technology capable of diffusing across the physical economy faster than states can track, regulate, and absorb. The problem, then, is not automation alone. It is the interaction of limited visibility, institutional mismatch, and delayed adjustment.

\section{A Compliance Architecture for the Autonomy Economy}
\label{sec:arch}

If the core problem is governance lag, then the next question is what kind of governance architecture can reduce it. That question cannot be answered by treating embodied AI as an isolated robotics issue. As argued in the Introduction, embodied AI is part of the autonomy economy, where autonomous capability scales across sectors through cumulative improvements in models, software, data, and robotic platforms. The policy challenge therefore extends beyond individual robots or single-product regulation. It is about governing how autonomous capability is deployed, controlled, and diffused across the physical economy.

A compliance architecture for the autonomy economy should be understood as a direct response to the three lags identified in the previous section. Observational lag requires institutions that can see deployment early and clearly. Institutional lag requires institutions that can assign responsibility and act on what they observe. Distributive lag requires institutions that can absorb the resulting gains and losses before they harden into wider instability. The four requirements below are therefore not separate policy ideas, but a connected architecture. Deployment visibility primarily reduces observational lag; stack-level accountability addresses institutional lag; trigger-based adjustment converts visibility and accountability into timely intervention; and automatic distributional response reduces distributive lag by stabilizing affected workers and regions before disruption becomes entrenched. These relationships are primary rather than exclusive, and several elements may need to operate in parallel. Table~\ref{tab:policy_mapping} summarizes the four policy recommendations proposed in this article, the governance lag each one addresses, and the rationale for doing so.

\begin{table}[t]
\caption{Policy recommendations and their relationship to the three governance lags}
\label{tab:policy_mapping}
\centering
\footnotesize
\begin{tabular}{p{0.34\linewidth} p{0.22\linewidth} p{0.28\linewidth}}
\toprule
\textbf{Policy recommendation} & \textbf{Governance lag addressed} & \textbf{Concise rationale} \\
\midrule
Standardized deployment reporting & Observational lag & Make deployment and labor effects visible early. \\
\addlinespace
Stack-level accountability rules & Institutional lag & Assign responsibility where control resides. \\
\addlinespace
Trigger-based adjustment mechanisms & Observational + Institutional lag & Turn visibility into timely intervention. \\
\addlinespace
Automatic adjustment funds & Distributive lag & Absorb uneven gains and losses early. \\
\bottomrule
\end{tabular}
\end{table}

The first requirement is \textit{deployment visibility}, which directly addresses observational lag. Governments cannot govern what they cannot see. In the autonomy economy, the relevant unit is not simply the number of robots installed, but the pattern of deployment: where autonomous systems are introduced, which tasks they replace or reorganize, how workflows change, and what effects follow for employment, wages, hours, and regional labor demand. Existing robot statistics remain useful, but they are not enough~\cite{worldrobotics2025industrial,ifr2025china2million}. Large deployers, systems integrators, and platform operators in high-exposure sectors should therefore be required to submit standardized, periodic disclosures on embodied-AI deployment and workforce impact. Without such reporting, observational lag will remain built into the governance system.

The reporting layer should be technically implementable. A minimum schema would identify the deployed system and model version, the responsible provider and integrator, the deployment site, the affected task categories, the operating hours of the autonomous system, material workflow changes, and aggregate workforce indicators. Common telemetry and reporting schemas would allow sector-level aggregation, while access controls and aggregation rules would protect commercially sensitive and worker-level data.

The second requirement is \textit{stack-level accountability}, which addresses institutional lag. Seeing deployment is not enough if institutions cannot determine where control resides and who must bear responsibility. The autonomy economy is governed through a layered stack of compute, models, platforms, robotic hardware, systems integration, and deployment channels. Accountability therefore cannot rest only on the final deploying firm. Upstream actors shape downstream outcomes. Model providers influence capability and behavior. Platform providers shape interoperability, switching costs, and market access. Integrators determine how autonomous systems are embedded into real workflows. Deployers determine where labor-market effects are ultimately realized. A workable compliance regime must therefore assign obligations across the stack, including disclosure, interoperability, incident notification, and review of vertically integrated control. This is especially important because AI infrastructure markets already display concentration, barriers to entry, and strategic control over key inputs~\cite{oecd2025aiinfrastructure,cma2024foundationmodels}. In the autonomy economy, responsibility must follow control.

Technically, such attribution requires provenance sufficient to connect model versions, platform components, integrator configurations, and deployment outcomes. Partial analogues exist in software bills of materials, model registries, and incident-reporting systems used in safety-critical sectors. The objective is not to impose undifferentiated joint liability across the stack. It is to assign each actor obligations proportionate to the decisions and controls it exercises, while preserving a traceable chain from upstream capability choices to downstream deployment.

The third requirement is \textit{trigger-based adjustment}. This is the bridge between observation and response. Its purpose is to ensure that visibility and accountability lead to timely institutional action rather than remain passive information.

Triggers can function only after a minimum visibility layer has been established. During the transition, governments should rely on coarse but observable proxies, such as major deployment announcements, capital expenditure on autonomous systems, robot-density changes, workforce-reduction notices, and regional wage or hours trends. As standardized reporting improves, these proxies can be replaced by more direct measures of task substitution and workflow change. Early triggers will be imperfect, but transparent, revisable thresholds are preferable to waiting for fully visible ex post disruption.

Governance cannot wait for crisis management after labor-market disruption has already spread. Adjustment mechanisms should activate when predefined thresholds are crossed, such as rapid increases in sectoral deployment intensity, concentration of adoption, measurable task substitution, or declines in local wages and hours. Once triggered, the policy response should be concrete: advance worker notification, firm co-financed retraining, targeted labor protections, and expedited competition review where concentrated platform control is implicated. The objective is not to block deployment, but to reduce the gap between technological change and institutional response.

The fourth requirement is \textit{automatic distributional response}, which addresses distributive lag directly. Its purpose is to reduce the gap between where gains concentrate and where losses are felt. The autonomy economy may generate productivity gains quickly, but gains and losses will not arrive at the same speed or in the same places. Gains are likely to concentrate first among the owners of capital-intensive systems, platforms, and deployment channels, while losses spread through bargaining power, hours, earnings, and regional labor-market exposure. Existing social-protection systems are poorly matched to this pattern because they often respond after formal job loss and have incomplete coverage~\cite{worldbank2025statesocialprotection}. Governments should therefore establish automatic adjustment funds for high-exposure sectors, financed through employer contributions or deployment-linked mechanisms, to support wage insurance, retraining, transition assistance, and local stabilization. If concentration begins early, adjustment must also begin early.

\textbf{Illustrative warehouse pilot.}
Consider a warehouse operator deploying autonomous mobile robots across multiple facilities. A minimum reporting schema could record the number and operating hours of deployed systems, affected task categories, worker hours by task category, workforce changes, safety incidents, and the share of workflow controlled by a common platform or integrator. A first-pass review trigger might be activated when autonomous-system operating hours rise by more than 20\% within a year while worker hours in affected task categories fall by more than 10\%, or when one platform controls more than half of automated workflow capacity in a defined regional market. These thresholds are illustrative rather than empirically validated. Their purpose is to show how deployment visibility can be converted into a reviewable institutional action rather than to prescribe universal numerical rules.

\textbf{Implementation constraints.}
Implementation will involve trade-offs. Embodied-AI supply chains are transnational, so disclosure obligations should attach primarily to market access and the jurisdiction of deployment rather than depend only on the location of upstream providers. Reported indicators may be gamed, which argues for auditable telemetry, third-party verification, and periodic revision of thresholds. Compliance costs should be scaled by deployment size, sectoral exposure, and risk so that small experiments are not treated like economy-wide rollouts. Deployment-linked financing for adjustment funds should likewise be calibrated to avoid discouraging beneficial adoption. Finally, the proposed architecture should complement rather than duplicate product-safety, AI-risk-management, labor, and competition regimes. Its distinctive role is to connect deployment data, stack-level responsibility, and labor-market adjustment across instruments that currently operate separately.

The interaction among these requirements is the core of the framework. Visibility without accountability produces information without action. Accountability without triggers produces responsibility without timely response. Adjustment without distributional support leaves the deepest social consequences to emerge after the fact. Visibility is a prerequisite for well-calibrated triggers, but accountability, reporting, and distributional preparation can develop in parallel. A workable compliance architecture must connect all four elements while allowing their sequencing to vary by sector and stage of deployment. Otherwise, the autonomy economy may scale faster than the institutions meant to govern it.

\section{Conclusion}
\label{sec:conc}

This article has framed governance lag as an embodied-AI-specific extension of the established pacing problem. Industrial-robot statistics show the scale of the existing physical-automation base, but direct evidence for general-model embodied-AI deployment remains limited; the article therefore advances a forward-looking governance argument rather than claiming that such systems have already diffused economy-wide. Embodied AI nevertheless creates distinctive pressures through potential economies of scale, task-level reorganization that may escape job-level statistics, and the separation of upstream control from downstream social impact. We organized these pressures into three mutually reinforcing lags, observational, institutional, and distributive, and proposed a practical way to assess them through reporting latency, institutional response time, distributional adjustment time, and coverage indicators. We then mapped the lags to four policy instruments: deployment visibility, stack-level accountability, trigger-based adjustment, and automatic distributional response.

The next step is empirical and institutional. Sector-specific pilots should test common reporting schemas, provenance mechanisms, lag indicators, and revisable trigger thresholds. Comparative research should examine how the EU AI Act as amended in 2026, machinery-safety law, robotics standards, AI management systems, and the NIST AI RMF can support these pilots and where additional coordination is required. The framework should also be evaluated against implementation risks, including jurisdictional fragmentation, metric gaming, compliance cost, and unintended effects on beneficial adoption. Governance lag will not be eliminated by a single rule. It can, however, be bounded and reduced if deployment becomes observable, responsibility follows control, and adjustment begins before disruption is fully entrenched.

%---------------------------------------------
% Author biographies
%---------------------------------------------
\begin{IEEEbiographynophoto}{Dr.\ Shaoshan Liu} is Director of Embodied AI at the Shenzhen Institute of Artificial Intelligence and Robotics for Society (AIRS). His research focuses on embodied AI, computer architecture, and public policy. He received his PhD from UC Irvine and MPA from Harvard Kennedy School. He is a Senior Member of IEEE. Contact him at \texttt{shaoshanliu@cuhk.edu.cn}
\end{IEEEbiographynophoto}

% Reviewer 2 copy-edit reminder: in references.bib, ensure the Marchant chapter DOI
% is encoded so that the rendered DOI is 10.1007/978-94-007-1356-7_2.
\bibliographystyle{ieeetr}
\bibliography{references}
\end{document}